\documentclass[%
reprint,
amsmath,amssymb,
aps,
]{revtex4-1}

\usepackage{graphicx}
\usepackage{dcolumn}
\usepackage{bm}
\usepackage{textcomp}
\usepackage{gensymb}
\usepackage{amsmath}
\usepackage{sidecap}
\usepackage{color}

\begin{document}

	\setlength{\abovedisplayskip}{10pt}
	\setlength{\belowdisplayskip}{10pt}	
	
	\title{Diffusion-related lifetime of indirect excitons in diamond}

	\author{K. Konishi$ ^{1} $}
	\author{I. Akimoto$ ^{2} $}
	\author{J. Isberg$ ^{3} $}
	\author{N. Naka$ ^{1} $}	
	\affiliation{%
		$ ^{1} $Department of Physics, Kyoto University, Kitshirakawa-Oiwake-cho, Sakyo-ku, Kyoto 606-8502, Japan}
	\affiliation{%
		$ ^{2} $Department of Materials Science and Chemistry, Wakayama University, Wakayama 640-8510, Japan}
	\affiliation{%
		$ ^{3} $Department of Electrical Engineering, Uppsala University, Box 65, S-751 03 Uppsala, Sweden}%
	
	\date{\today}
	\begin{abstract}
We investigate the lifetime of indirect excitons in extremely high purity diamond grown by the chemical vapor deposition method. 
A clear correlation is found between the lifetime and small strain (magnitude $< $10$^{-4}$) assessed using birefringence, thanks to the effective absence of impurity traps. 
A surface recombination model, extended for nonradiative recombination at dislocations 
due to exciton diffusion, is proposed to explain the temperature-dependent lifetime. 
Based on the derived radiative lifetime, our model enables the prediction of lifetimes at any temperature 
as well as the highest achievable internal quantum efficiency of exciton luminescence in diamond, which can generally be applicable to a wide range of materials with high exciton binding energies.

		\begin{description}
			\item[DOI]
			10.1103/			
		\end{description}
	\end{abstract}
	
	\pacs{Valid PACS appear here}
	\maketitle
	
	
{\it Introduction -}	
The optical properties of wide bandgap semiconductors are generally characterized by excitons, i.e., pairs of an electron and a hole bound
by the Coulomb force. 
This is contrary to conventional semiconductors, in which excitons have an effect at cryogenic temperatures only.
Among various wide bandgap materials,
diamond is a particularly robust solid-state platform for visible light emission from nitrogen-vacancy centers acting as a 
single photon source \cite{NV} and for deep-ultraviolet (DUV) emission from excitons of intrinsic origin, 
which has been receiving increasing interest owing to sterilization applications effective at 
a shorter wavelength (235 nm) \cite{Makino} than the limit of GaN-based blue light-emitting diodes (LEDs).
In diamond, excitons are formed across the indirect bandgap (5.5 eV) and, due to their large binding energy, 
dominate the DUV emission up to room temperature.

So far, dynamics studies of optically excited carriers have mostly been performed 
using photoinduced transient grating (PITG) and time-resolved photoluminescence (TRPL) methods.
Generally, transient carriers and excitons are subject to various complex pathways, such as 
trapping, diffusion, surface recombination \cite{PRAppl},
and the radiative and nonradiative bulk recombinations. 
In 1996, a pioneering TRPL report by Takiyama {\it et al.} \cite{Takiyama} clarified the temperature dependence of the bulk lifetime in moderate quality diamonds.
The reductions of the lifetime at low and high temperatures were attributed to the trapping at impurities with
the formation of bound excitons and to the exciton ionization, respectively.
Although a reasonably long radiative lifetime, greater than 250 ns, 
was estimated, the derived activation energy for the bound excitons was in disagreement with the spectroscopic data, 
which has still not been clarified. 

Over the two decades following the Takiyama report, with the advance of chemical-vapor-deposition (CVD) growth methods \cite{Balmer} the purity of synthetic diamond has been dramatically improved. 
This motivated the review of the long-lasting question of using higher purity diamonds.
Using PITG methods \cite{Kozak_Temperature_2014,Scajev_Excitation_2017} on CVD-grown high purity diamonds, 
the diffusion coefficient and surface recombination velocity were extracted.
The exceptionally high diffusivity was found using the TRPL method \cite{MorimotoPRB}; however, the temperature dependence of the lifetime was 
discussed only qualitatively, based on a three-level model for the free exciton, bound exciton, and free carriers \cite{Morimoto_Exciton_2016}.
Thus, the detailed mechanisms determining the decay of indirect excitons still have to be fully established.	

In this study, we systematically evaluated the exciton lifetime in different diamond samples, and determine
the influence of strain on the 
recombination rates.
For this purpose, we used ultrapure diamonds to ensure that the exciton decay was not affected by trapping at impurities.
We combined information gained from different techniques, such as the exciton diffusion coefficient by the TRPL, strain distribution using birefringence imaging, and carrier mobility measured by the time-of-flight (ToF) method.
We provide the first comprehensive explanation of  the temperature dependent exciton decay in diamond.
Our analysis reveals that the exciton decay is mostly determined by the arrival rate at a dislocation due to diffusion 
from 4 K to 300 K, while it is also affected by the bulk radiative lifetime at intermediate temperatures. 
This result is useful for optimizing the design of diamond devices with the exciton lifetime and luminescence quantum efficiency
determined quantitatively. 
Furthermore, the simple model proposed in this study can conveniently be extended to a wide range of excitonic materials, such as atomically-thin semiconductors \cite{mono}, nitrides, and perovskites, when high-quality samples are synthesized.

	\begin{figure*}[tb]
	\centering
	\includegraphics[width=17cm,clip,bb=15 185 832 357]{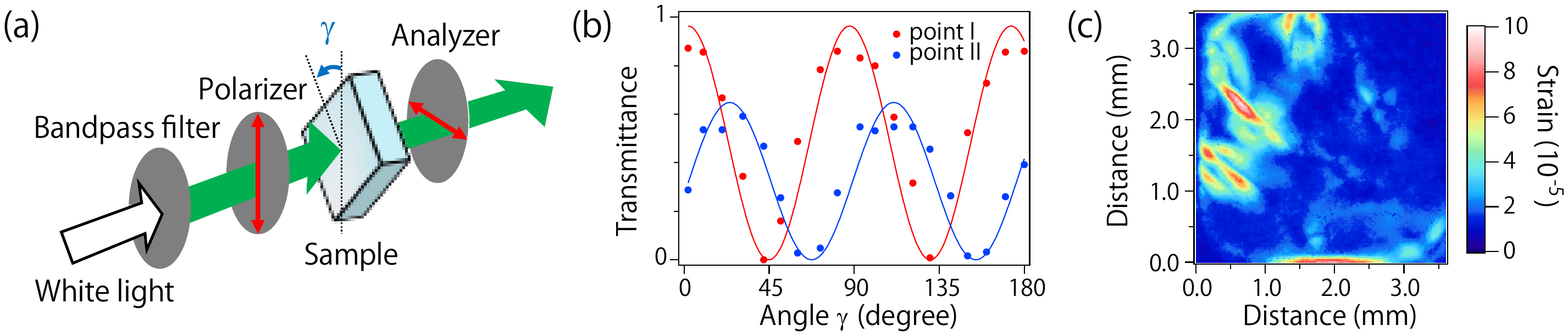}
	\caption{(a) Schematic of the configuration taking birefringence images using a polarization microscope. (b) Dependence
	of the transmitted light intensity on the rotation angle at two points (I, II) on sample 3. (c) Map of the strain magnitudes obtained
	from the analysis of the birefringence images of sample 2.}
	\label{fig:fig1}
	\end{figure*}

{\it Sample characterization -}
We used three single-crystal CVD diamonds (samples 1, 2, and 3) of the highest currently available purity, supplied by Element Six. 
Sample 2 is a commercially available electric-grade sample. 
The nitrogen impurity concentrations of sample 2 and 3, measured by electron paramagnetic resonance \cite{Shimomura}, were 0.07 $ \pm $ 0.02 ppb and 0.05 $ \pm $ 0.03 ppb, respectively. In all three samples the boron concentrations were lower than 0.2 ppb.
The dimensions of samples 1, 2, and 3 are $ 3.9\times 4.0\times 0.51 $ mm$^3$, $3.5\times3.6\times0.53$ mm$^3$, and $ 4.4\times4.5\times0.49 $ mm$^3$, respectively.

We quantified the strain distribution by taking birefringence images \cite{Pinto_Point_2011}. 
The experimental arrangement is shown in Fig. \ref{fig:fig1}(a). 
The images of the sample between crossed polarizers were recorded with the rotation angle $\gamma$ varied under a microscope.
The transmittance at each point can be expressed as
	\begin{equation}	
		T/T_{0}=\sin^{2}\left( \delta/2\right) \sin^{2}\left( 2\gamma\right),
		\label{eqS1}
	\end{equation}
where $ \delta $ is the phase difference between the two polarized components of the transmitted light. 

The measured $ \gamma $-dependence is shown in Fig. \ref{fig:fig1}(b) for two different points on sample 3. 
The curves of the best fit with Eq. (1), to extract $\delta$ \cite{note}, are plotted by lines. 
The change in the refractive index is then obtained as $ \varDelta n = (\lambda/2\pi d)\delta $, where $ \lambda $ is the incident light wavelength and $ d $ is the sample thickness. 
Using the refractive index value $ n=2.4 $ and the photoelastic constant $p= p_{11}-p_{12}=-0.3 $  for diamond \cite{Hounsome_Photoelastic_2006}, the magnitude of the strain can be calculated as $ \epsilon=\varDelta n/(n^{3}p) $ \cite{Pinto_Point_2011}.  
We used monochromatic light of $ \lambda=(560\pm5) $ nm 
instead of white-light illumination, by inserting a bandpass filter (Thorlabs, FB 560-10). 
This enabled the elimination of the ambiguity in $ \lambda $ in determining $ \varDelta n $, and the quantification of a subtle strain ($<10^{-4} $, which is too small to be seen as Raman shift \cite{Crisci}) with a spatial resolution of 10 $ \mu $m. 
The magnitude of the strain at each pixel on sample 2 is shown in color in Fig. \ref{fig:fig1}(c). 
We found that non-uniform strains were present in each sample.

The maximum value of the strain, $ \epsilon_{\rm max}  $, and the strain averaged over the respective sample areas, 
$ \epsilon_{\rm avg}$, were found to increase in the order of samples 1, 2, and 3 (Table I). 
The strain at the excitation spot, $\epsilon_{\rm spot}$, for TRPL measurements is also listed in Table I, for the centers of 
samples $1-3$ (spots $1-3$) and the off-center of sample 3 (spot 3').

\begin{table}[t]
\setlength{\tabcolsep}{12pt}
\renewcommand{\arraystretch}{1.2}
\begin{tabular}{c|c|c|c|c}
\hline
Sample/spot &$1/1$&$2/2$&$3/3$&$3/3'$ \\  \hline\hline
$\epsilon_{\rm max}$ (10$^{-5}$) 
&$8.9$&$9.6$&\multicolumn{2}{c}{19}\\ \hline
$\epsilon_{\rm avg}$ (10$^{-5}$)
&$1.6$&$2.0$&\multicolumn{2}{c}{4.0}\\ \hline
$\epsilon_{\rm spot}$ (10$^{-5}$)
&$1.3$&$2.0$&$9.0$&3.7\\ \hline
\end{tabular}
\caption{Strain magnitude assessed for samples $1-3$.}
\end{table}
	
\begin{figure}[t]
\centering
\includegraphics[width=8.5cm,clip,bb=90 30 785 518]{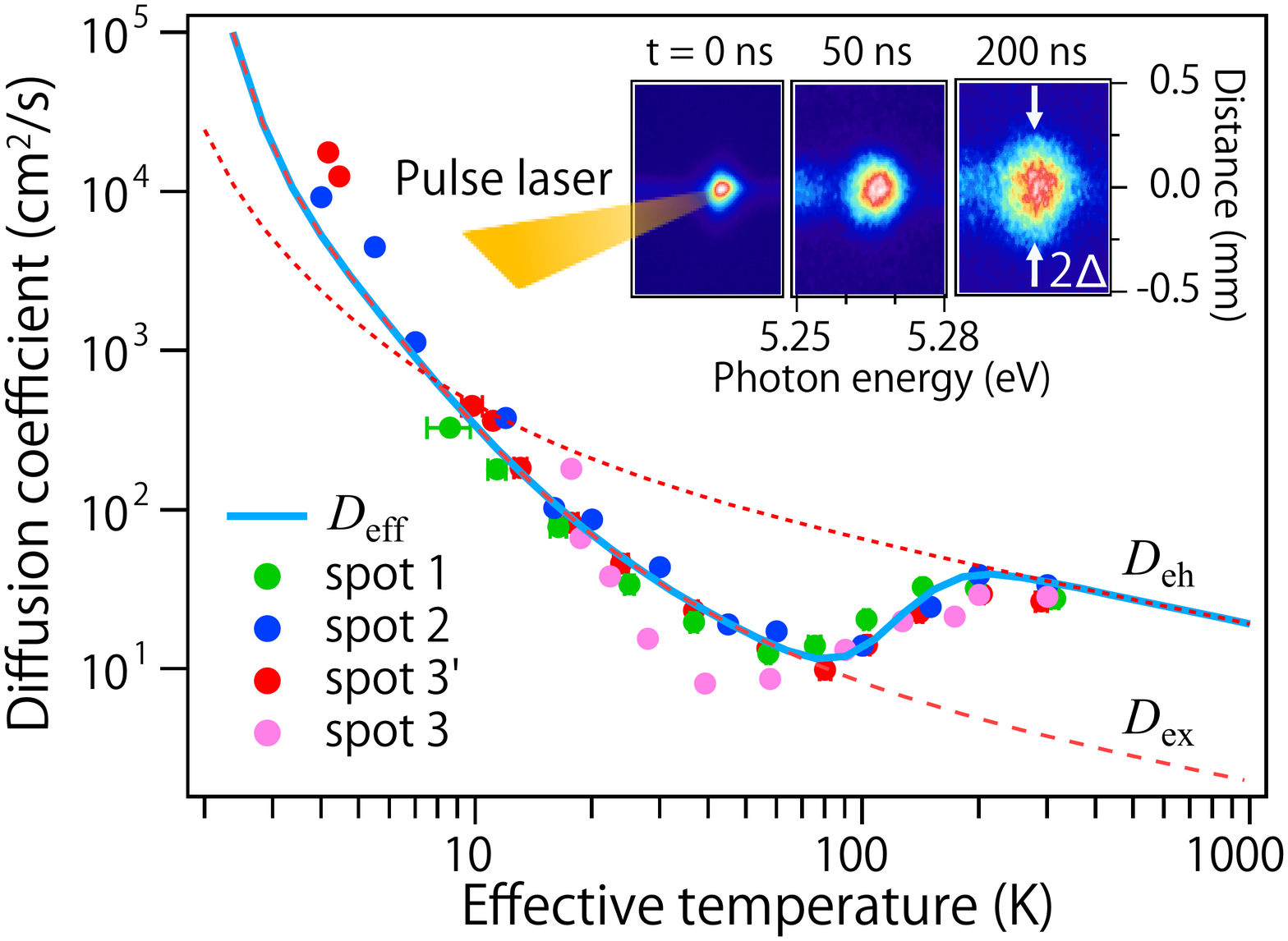}
\caption{Exciton diffusion coefficients measured at spots $1-3$ and 3'. Data points for spot 2 are taken from Ref. \cite{MorimotoPRB}.
Lines represent calculated values (see text). Inset shows spectral images of the exciton cloud at different delay times, obtained on sample 1 at 7 K.}
\label{fig:fig2}
\end{figure}

{\it Modeling the exciton diffusion -}
The exciton diffusion was measured using the space-resolved TRPL method reported in Ref. \cite{MorimotoPRB}. 
As shown in the inset of Fig. \ref{fig:fig2}, the pulsed laser beam 
of 2.5 ns duration was focused into a small spot (diameter of $\sim 30$ $\mu$m) on the sample to create excitons. 
The spatial expansion was recorded by imaging the exciton luminescence cloud through a monochromator 
onto a gated charge-coupled device (CCD) camera (see Supplemental Material). 
The time evolutions shown in the inset were analyzed as a function of the delay time $t$ following photoexcitation.  
The diffusion coefficient $D$ was obtained using the formula
$D=\Delta^2/(4 t)$ derived from the diffusion equation, where $\Delta$ is the half width of the exciton cloud.
The measurements were repeated at various temperatures and on different samples.
The results at four spot positions are plotted in Fig. \ref{fig:fig2} as a function of the effective temperature of the excitons
\cite{Hazama}.
No significant difference was found between the diffusion coefficients at spots 1, 2, and 3', which indicated that a small 
amount of strain ($\epsilon_{\rm spot} < $ $4\times10^{-5}$) did not affect the diffusion coefficient. 

The dashed line in Fig. \ref{fig:fig2} represents the diffusion coefficient $D_{\rm ex}$ obtained using the calculated rate of acoustic phonon scattering \cite{MorimotoPRB}, which is in good agreement with the data points up to 80 K. 
As indicated in Ref. \cite{MorimotoPRB}, the deviation above 80 K results from the thermal ionization of excitons into electrons and holes with higher mobility. 
In this paper, we use the ToF mobility $\mu_{i}$ reported in the literature \cite{ToF} and the corresponding diffusion coefficient $D_{i}=\mu_{i} k_BT/q$ for electrons ($i$=e) and holes ($i$=h) \cite{note1}.
As the in-plane electron diffusion is considered an average effect of the propagation of hot electrons (with transverse effective mass $m_t=0.28m_0$, where $m_0$ is 
the free electron mass) and cool electrons (with longitudinal effective mass $m_l=1.56 m_0$), we use the mean value based on the conductivity effective mass, $m_{\rm c}=3/(2/m_{\rm  t}+1/m_{\rm l})$.  
Considering that electrons and holes after diffusing around as free carriers then recombine into excitons again, 
we assume that the product of electron and hole densities propagates with an effective diffusion coefficient of 
$1/D_{\rm eh}=1/D_{\rm e}+1/D_{\rm h}$.
In Fig. \ref{fig:fig2}, $D_{\rm eh}$ is indicated by a dotted line.
The weighted sum of exciton and carrier diffusivity, $D_{\rm eff}=f(T) D_{\rm ex}+(1-f(T))D_{\rm  eh}$, 
is indicated by the thick line, where the exciton fraction $f(T)$ is considered according to the law of mass action. In the mass action law,
a total carrier density of $n=1\times 10^{15}$ cm$^{-3}$, the exciton reduced mass of $0.19 m_0$, and the exciton binding energy of 94 meV were used \cite{Ichii}.
The calculated line is in good agreement with the measured diffusion coefficient in the entire temperature range and it is used in the following analysis.

\begin{figure}[bth]
	\centering
	\includegraphics[width=8.5cm,clip,bb=85 118 390 487]{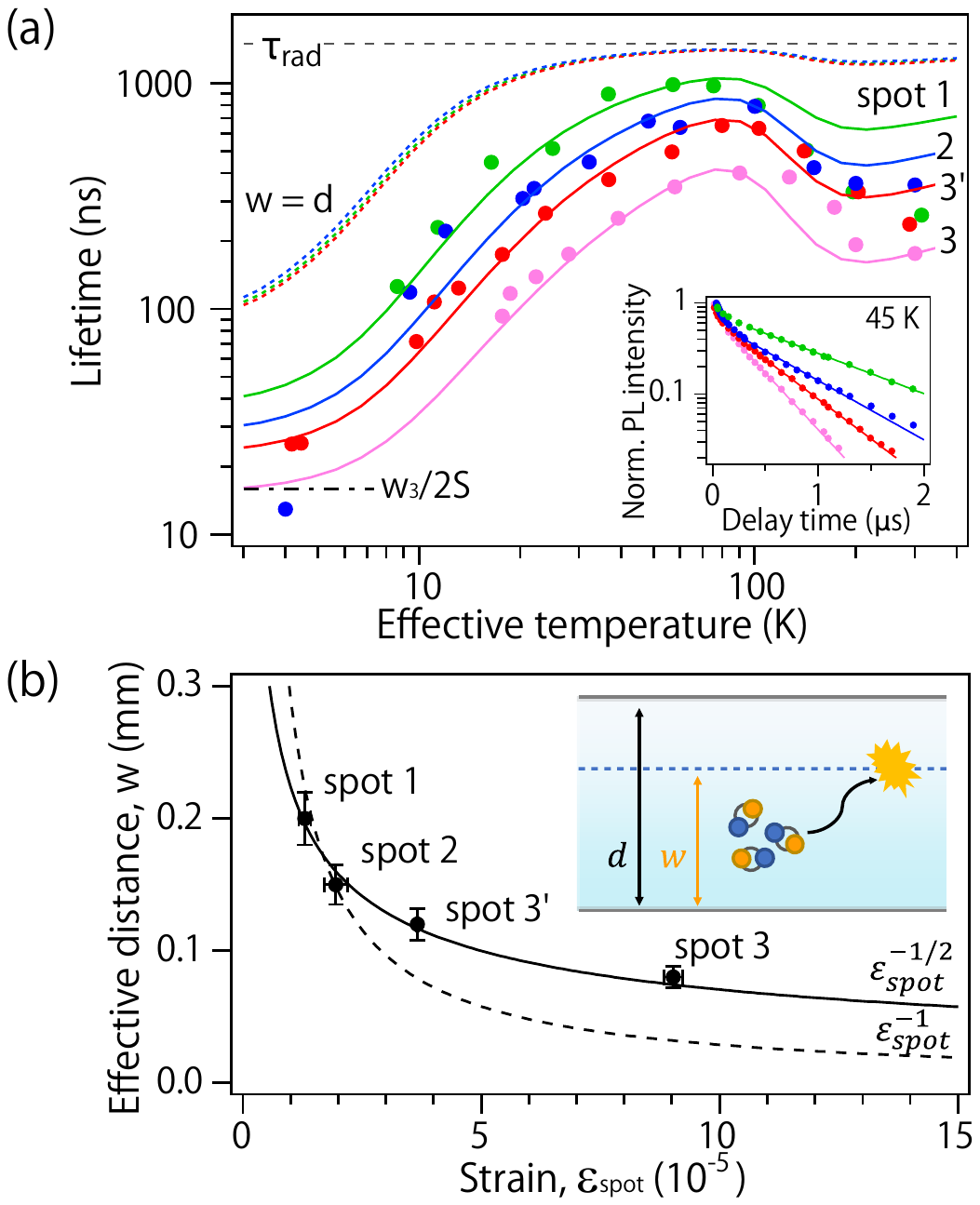}
	\caption{
	(a) Exciton lifetimes measured at different spot positions. Solid lines show $\tau_{\rm net}$ with $w$ adjusted, while dotted lines were calculated with $w=d$. 
	Data points for spot 2 are taken from Ref. \cite{Morimoto_Exciton_2016}. Inset shows PL decay curves obtained at 45 K. 
	(b) Effective distance as a function of strain magnitude. Inset is a cross-sectional view of exciton diffusion and nonradiative recombination at a dislocation.}
	\label{fig:fig3}
	\end{figure}

{\it Strain effects on exciton lifetime -}	
A similar series of data as for diffusion enables the extraction of exciton decay times \cite{Morimoto_Exciton_2016}.
The signal was spatially integrated over the cloud and spectrally integrated over the exciton luminescence 
for each delay time. The total photoluminescence (PL) intensity was analyzed assuming an exponential decay 
as shown in the inset of Fig. 3(a).
The decay times obtained at spots $1-3$ and 3' are plotted as a function of temperature in Fig. \ref{fig:fig3}(a). 
The decay time was shortest at spot 3 where the largest strain was observed.
The decay time also exhibited a noticeable temperature dependence.

In previous studies on moderate purity diamonds \cite{Takiyama},
the decay time was modeled in the range $75- 300$ K by considering nonradiative processes in the bulk lifetime, i.e., 
impurity trapping and 
exciton ionization. 
Contrary to this, we found that the impurity trapping occurred significantly slower than the radiative decay in the present samples, based on the upper limits of the impurity concentrations and capture cross sections \cite{Shimomura}.
The possibility of exciton two-body annihilation was also excluded, considering the low exciton density ($< 10^{16}$ cm$^{-3}$).
Therefore, although the bulk lifetime generally includes nonradiative and radiative contributions,
the radiative lifetime is crucial in our effectively impurity-free environment.
Diamond is an indirect bandgap semiconductor, and thus exhibits slow bulk radiative recombination, with 
reported radiative lifetimes $\tau_{\rm rad}$ of 2.3 $\mu$s \cite{Fujii} or 750 ns \cite{Kozak2013}.

It is known that for various semiconductors, the exciton lifetime can be described by the sum of the rates due to bulk and 
surface decay, as \cite{textbook}
\begin{equation}
\tau_{\rm net}=\left (\frac{1}{\tau_{\rm bulk}}+\frac{1}{\tau_{\rm surf}} \right)^{-1}.
\label{taunet}
\end{equation}
Due to the slow bulk decay in diamond, the dominant mechanism limiting the net lifetime is considered as the surface lifetime $\tau_{\rm surf}$. 
The surface lifetime consists of the rate of recombination at the surfaces and the arrival rate of the excitons diffusing to the surfaces.
When the sample has two equal sides, the surface lifetime is given by \cite{Sproul}
\begin{equation}
\tau_{\rm surf}=\frac{d}{2S}+\frac{1}{D} \frac{d^2}{\pi^2},
\end{equation}
where $S$ is the surface recombination velocity.
The first and second terms are obtained
under the limits of $S=0$ and $S=\infty$, respectively, for the fundamental-mode solution of the diffusion equation.  

By modifying the above formula,  
we used the relation: 
\begin{equation}
\tau_{\rm disl}=\frac{w}{2S}+\frac{1}{D_{\rm eff}} \frac{w^2}{\pi^2}.
\label{tausurfw}
\end{equation}
Here, we replaced the sample thickness $d$ by an effective distance $w$ to a dislocation resulting in the strain, to consider the diffusion to and recombination at the dislocation [see the inset of Fig. \ref{fig:fig3}(b)]. 
The solid lines in Fig. \ref{fig:fig3}(a) represent $\tau_{\rm net}$ calculated by using $\tau_{\rm disl}$ instead of $\tau_{\rm surf}$ in Eq. (\ref{taunet}). 
We used $D_{\rm eff}$ as shown in Fig. \ref{fig:fig2} and adjusted the three parameter values: $S$, $w$, and the bulk lifetime approximated by a radiative lifetime ($\tau_{\rm bulk}=\tau_{\rm rad})$. We obtained reasonable fitting results with $S=0.25\times10^6$ cm/s and $1.4 \leq \tau_{\rm rad} \leq1.8$ $\mu$s, and here we use the value of $\tau_{\rm rad}=1.5$ $\mu$s.
Our $S$ value is slightly smaller than the literature value \cite{Kozak2012}.
The measured points are nearly perfectly reproduced at $w=0.20$, $0.15$, $0.08$, and  $0.12$ mm for spots $1-3$ and 3', respectively.
The contribution of the surface recombination velocity [the first term in Eq. (\ref{tausurfw})] is shown for spot 3 
by the dash-dotted line, which limits the low-temperature lifetimes only.

A smaller $w$ value was obtained at a spot with higher strain. The dependence of $w$ on $\epsilon_{\rm spot}$ is shown in Fig. \ref{fig:fig3}(b). 
We empirically found that the data points followed the relationship $w\propto 1/\sqrt{\epsilon_{\rm spot}}$ (solid line) rather than $w\propto 1/\epsilon_{\rm spot}$ (dashed line).
Under the presence of dislocations at density $N_d$, the arrival time can be expressed as $t=4/(D_{\rm eff} \pi^3 N_d)$ \cite{YamaguchiJAP,SiegAPL}. Comparing this with the second term of Eq. (4) results in $N_d=4/(\pi w^2)=3200$, 5700, 16000, and 8900 cm$^{-2}$ for spots $1-3$ and 3'.
These values are reasonable considering the state-of-the-art CVD growth with low defect densities \cite{Balmer}.

Finally, the dotted lines in Fig. 3(a) represent predictions when $w$ is set equal to the sample thickness of $d\simeq 0.5$ mm.
For this defect-free case, the exciton decay at intermediate temperatures is limited by the bulk radiative lifetime 
($\tau_{\rm rad} < \tau_{\rm disl})$; thus, a high internal quantum efficiency of 
$\eta=(1+\tau_{\rm rad}/\tau_{\rm nrad})^{-1} \simeq (1+\tau_{\rm rad}/\tau_{\rm disl})^{-1}$ \cite{PRAppl,Kozak_Temperature_2014,Fujii} can be achieved. 
The solid lines in Fig. 4(a) indicate $\eta$ calculated using $\tau_{\rm disl}$ given by Eq. (4). The highest quantum efficiency is expected at 100 K when the diffusion coefficient nearly takes its minimum value. 
This was observed in the exciton luminescence measured in the time-integrated regime [Fig. 4(b)], and the total PL intensity 
obtained by spectrally integrating these spectra, as indicated by dots in Fig. 4(a) (see Supplemental Material for details).
The drop in the exciton luminescence at high temperature is illustrated by
the dashed lines, representing $\eta$ multiplied by the exciton fraction $f(T)$ according to the mass action law \cite{Ichii}.
For temperatures below 80 K, the excitons undergo transition to electron-hole droplets
whose contribution is difficult to include in the plot due to their very fast nonradiative decay by Auger recombination \cite{Shimano}.

The plot indicates that excitons are mostly ionized into free carriers at room temperature \cite{hBN}, which migrate faster than excitons, resulting in a decrease in the lifetime due to the faster diffusion to a dislocation. 
This decrease can be prevented by increasing the total carrier density and thus 
increasing $f(T)$ and reducing $D_{\rm eff}$ (dash-dotted line).
Therefore, to achieve efficient excitonic luminescence from diamond LEDs high-current injection is required.
It should be emphasized that the exciton decay is limited by the diffusion-related term for a wide temperature range. Consequently,
understanding the exciton diffusion is relevant for determining the exciton lifetime and the luminescence
quantum efficiency, which is crucial for the electronic and optoelectronic applications of diamond for efficient devices.

\begin{figure}[t]
\centering
\includegraphics[width=8.5cm,clip,bb=10 40 830 525]{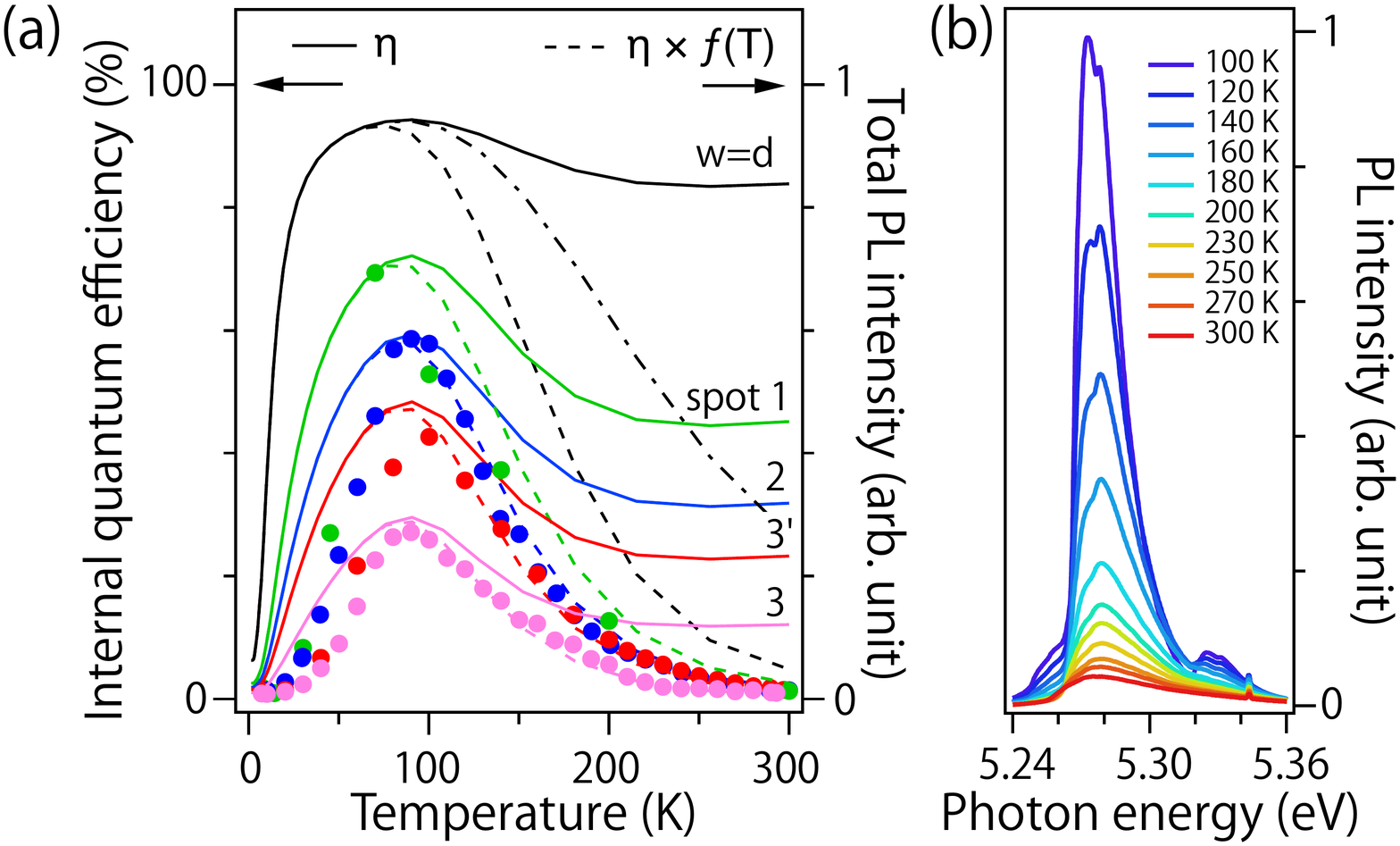}
\caption{(a) Solid lines: internal quantum efficiency $\eta$ of excitons calculated at $w=0.20$ mm (green), $0.15$ mm (blue), $0.12$ mm (red), $0.08$ mm (pink), and $0.5$ mm (black). Dashed and dash-dotted lines: $\eta$ multiplied by the exciton fraction $f(T)$ assuming a total density of $n= 4\times 10^{15}$ cm$^{-3}$ and $n=4\times 10^{16}$ cm$^{-3}$, respectively. Dots: spectrally integrated PL intensity, corrected by factors ($1.0-1.9$) for a detection efficiency with slight daily variations.
(b) Exciton luminescence spectra on spot 3' obtained at different temperatures.}
\end{figure} 

{\it Conclusion -}		
We systematically analyzed the lifetimes of excitons obtained using TRPL measurements in the range of $4-300$ K in extremely high purity CVD diamonds.
By excluding the effect of exciton trapping at impurities, we found a clear correlation between the exciton lifetime and the strain magnitude as small as $10^{-5}$.
The temperature dependence of the lifetime was successfully reproduced by the surface recombination model, extended for the case when the strain is acting as a recombination center according to the significant diffusion of excitons. 
The present study enables the prediction of the temperature-dependent exciton lifetime and luminescence quantum efficiency, 
which is useful in designing diamond-based devices such as LEDs and radiation detectors.
Furthermore, it generally facilitates a better understanding of the optical processes in a wide range of
semiconductors with high exciton binding energies.

\begin{acknowledgments}
This work was partially supported by JSPS KAKENHI (Grant Nos. 17H02910, 19K21849, and 15K05129),
JSPS bilateral project (No. 120209919), and the Swedish Research Council (Grant No. 2018-04154).
\end{acknowledgments}

\bibliographystyle{aps}

\begin{thebibliography}{99}
\bibitem{note}
The value of
$T_0$ was taken as the maximum transmission intensity in the parallel polarizers arrangement at a position where the strain was negligible, to consider the reflection at the sample surfaces.
The transmittance $T/T_0$ never reached unity at any points in the three samples, indicating that the condition $ 0< |\delta| < \pi$ was satisfied.
\end{thebibliography}


\clearpage
\onecolumngrid
\noindent
\vspace{12pt}
{\large Supplementary Material} \\ 
\noindent
\vspace{12pt}
{\large {\bf Diffusion-related lifetime of indirect excitons in diamond}}\\
K. Konish,$^1$ I. Akimoto,$^2$ J. Isberg,$^3$ Nobuko Naka$^1$\\
\normalfont\itshape
$^1$Department of Physics, Kyoto University, Kitashirakawa-Oiwake-cho, Sakyo-ku, Kyoto 606-8502, Japan \\
$^2$Department of Materials Science and Chemistry, Wakayama University, Wakayama 640-8510, Japan \\
$^3$Department of Electrical Engineering, Uppsala University, Box 65, S-751 03 Uppsala, Sweden

\vspace{24pt}

\twocolumngrid

\normalfont\rmfamily
\renewcommand{\thesection}{S\arabic{section}}
\def\thefigure{S\arabic{figure}}
\def\theequation{S\arabic{equation}} 
\setcounter{figure}{0}
\setcounter{equation}{0}

\section{Measurement of diffusion coefficient}

The method for extracting the diffusion coefficient of excitons in diamond was reported in our previous paper [1];
here, we summarize the main points only. 

The sample was mounted in a closed-cycle cryostat and the temperature $T_{\rm base}$ was measured by a silicon diode thermometer attached to the sample holder.  The excitation light source was an output from an optical parametric oscillator pumped by a YAG laser (Ekspla, NT242-2).
The excitation wavelength was 225 nm and the penetration length was approximately 500 $\mu$m at 100 K (see also Sect. S2), comparable to the sample thickness. 
The luminescence from the sample was imaged onto the entrance slit of the monochromator (Horiba Jobin Yvon, iHR550) by a pair of achromatic lenses with a magnification factor of $M=5/3$. 
Two exit ports were implemented in the monochromator; one was used for detection by a gated-ICCD camera (LaVision, PicoStarHR) synchronized to the trigger from the laser, the other for detection by a CCD camera (Andor, Newton DU940N) in a time-integrated regime. 
To ensure to collect the exciton luminescence only, the grating was set at the first-order diffraction mode
and the luminescence was spectrally resolved with the spatial information retained along the vertical slit of the monochromator. 
Figure S1(a) shows the spectral image at a 50 ns delay time obtained with sample 1 at 100 K.  

Figure S1(b) shows the luminescence spectrum obtained by vertically integrating the counts shown in Fig. S1(a).
The major peak is due to the recombination of the indirect exciton accompanied by the emission of a transverse-optical (TO) phonon.  
Intuitively, the width at temperature $T$ is 1.8 $k_BT$, where $k_B$ is the Boltzmann constant. 
More precisely, the effective temperature $T_{\rm eff}$ of excitons can be obtained by spectral fitting using the Maxwell-Boltzmann
distribution:
$I(E)=A \sqrt{E}\exp(-E/k_B T_{\rm eff})$, where $A$ is a normalization factor and $E$ is the photon energy measured from the edge 
[2]. We included four fine-structure exciton levels and the effect of spectral broadening due to the finite width of the entrance slit by convolution. The best fit function 
is indicated by the dashed line. The effective temperature $T_{\rm eff}=112$ K was slightly higher than $T_{\rm base}= 100$ K,
and the temperature increase was slightly different depending on samples. 

Figure S1(c) shows the spatial profiles at different delay times, obtained by spectrally integrating the luminescence image over the TO phonon assisted line. The profile was then fitted by a Gaussian function to extract the half width, $\Delta$. 
Figure S1(d) shows the squared half width as a function of the delay time $t$. A linear time dependence is expected from the diffusion equation, and the lines are fits to obtain the diffusion coefficient, $D=\Delta^2/(4t)$.

Similar measurements and analyses were performed for four different spot positions and at various temperatures,
providing the data points shown in Fig. 2 of the main text.  

\begin{figure}[t]
\centering
\includegraphics[width=8.5cm,clip,bb=30 30 795 540]{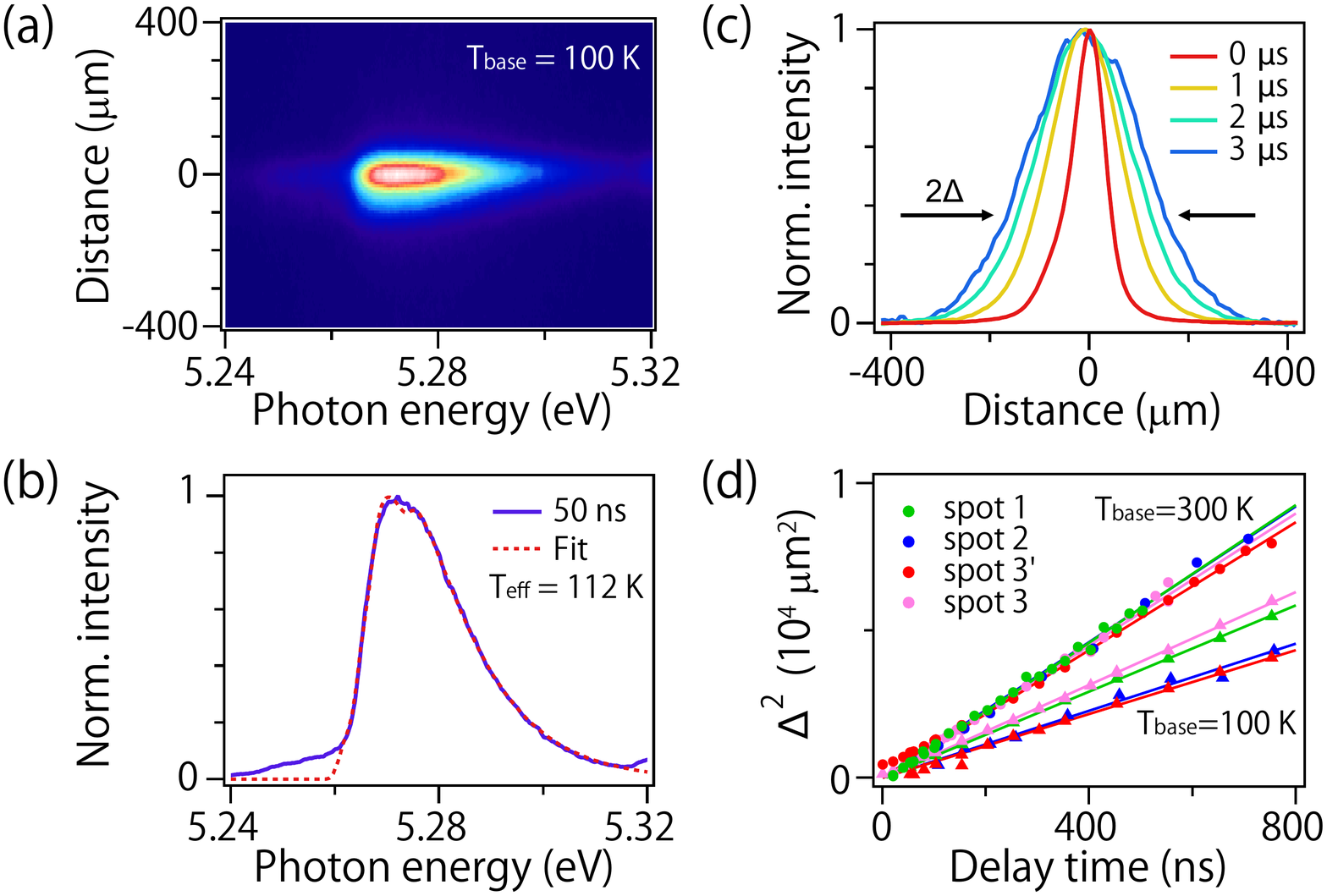}
\caption{(a) Spectral image of exciton luminescence in sample 1 at 100 K, gated at a 50 ns delay time. 
(b) Spectrum obtained by spatially (vertically) integrating the counts in the image in panel (a). 
(c) Spatial profile obtained by spectrally (horizontally) integrating the counts in the image in panel (a). The data 
obtained at various time delays are also shown. 
(d) Squared half width of the exciton cloud as a function of delay time, from four spot positions
at temperatures 100 K and 300 K. The initial broadening at $t=0$ has been subtracted
for clarity. The lines represent the best fits to extract $D$.}
\end{figure} 

\section{Measurement of luminescence intensity}

The exciton luminescence intensity measurements are fundamentally simple. In the following, we describe the effect of electron-hole droplets (EHDs) and certain corrections performed for the accurate evaluation of the intensity.

Figure S2(a) shows the time-integrated photoluminescence (PL) spectra obtained 
from spot 3' for temperature varied from 300 K to 7 K. 
The spectral range is wider than that shown in Fig. S1(b), and the transverse-acoustic (TA) and two-phonon assisted lines 
can be seen in addition to the TO phonon assisted line at 5.27 eV.
Both the spectral width and intensity changed drastically with the temperature.
The broadening of the line with increasing temperature was consistent with the thermal distribution of 1.8 $k_BT$, as explained above. 

\begin{figure}[t]
\centering
\includegraphics[width=8.5cm,clip,bb=75 20 760 575]{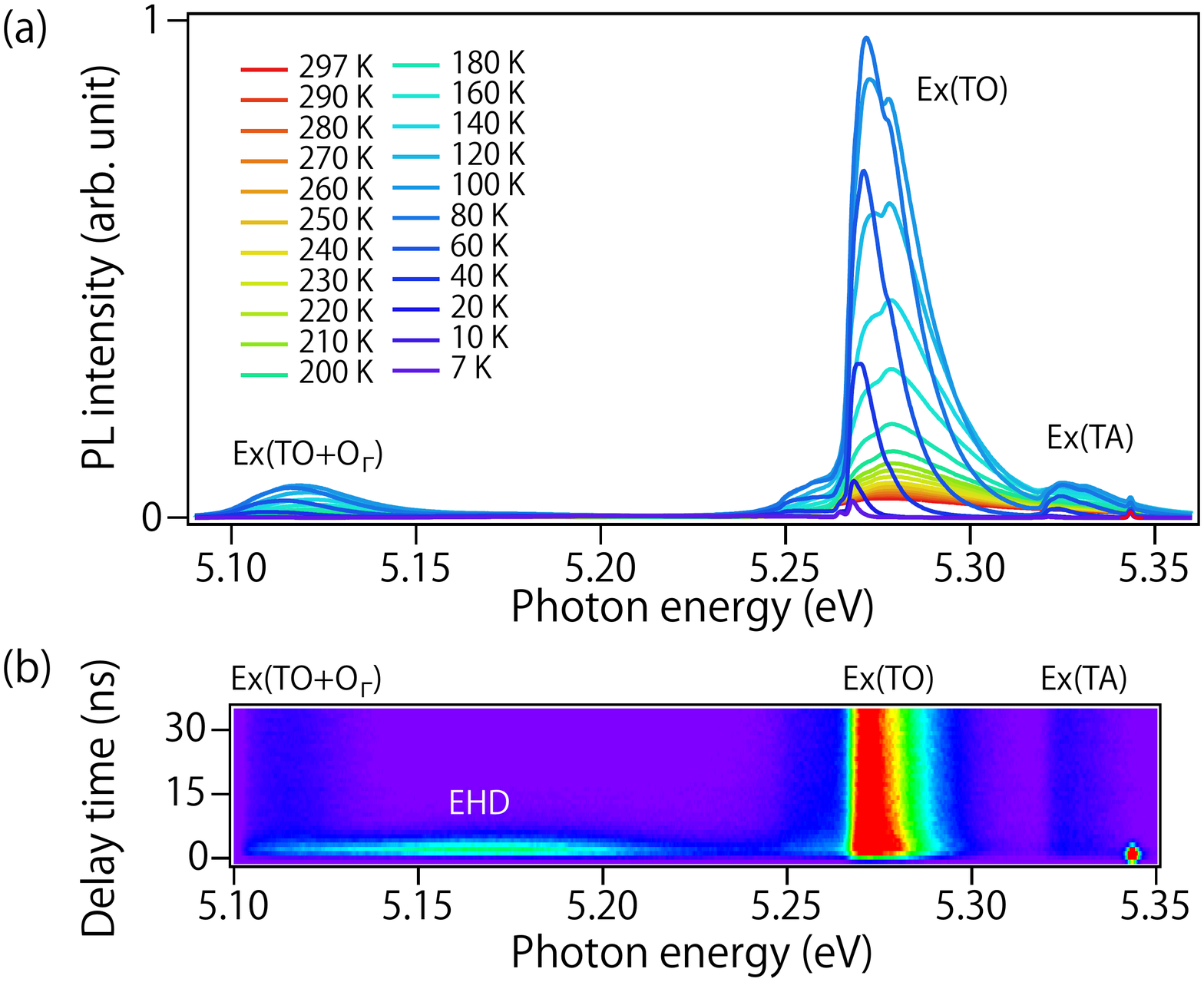}
\caption{(a) Time-integrated spectra of exciton luminescence on spot 3' at different temperatures. 
Label "Ex($\nu$)" denotes phonon-assisted recombination line of excitons (Ex) involving emission of a phonon of mode $\nu$.
(b) Time-resolved spectra of luminescence on spot 3' at 70 K, showing the EHD signal. The red spot at 5.34 eV is the Raman scattering signal. }
\end{figure} 

As mentioned in the main text, there were a few missing counts below 80 K due to the EHD formation. 
Figure S2(b) shows the time-resolved luminescence spectrum following photoexcitation. 
The broad structure seen in the range of $5.15- 5.20$ eV is due to the recombination of the EHD.
The decay is very fast $(< 1$ ns), and its contribution can hardly be seen in the spectra obtained 
in the time-integrated regime [Fig. S2(a)]. 
For a typical incident laser power of $\sim 100$ $\mu$W used in our experiments, the EHD signal was observed
below 80 K in the time-resolved regime. 

For the exciton luminescence efficiency, we spectrally integrated the time-integrated signal counts from 5.05 to 5.40 eV covering 
the three phonon-assisted lines. Then, we corrected the counts by the laser power absorbed by the sample. 
The laser wavelength was fixed at 225 nm for the whole temperature range, but the absorption coefficient is known to be
temperature dependent. As shown in Fig. S3(a), the absorption coefficient $\alpha$ increases exponentially 
with the temperature according to the increase in the phonon occupation number. 
The fraction of the absorbed photon over the total photon numbers was estimated using the Beer-Lambert law for the measured absorption coefficient $\alpha$ and the sample thickness $d$. The calculated fraction $g(T)=(1-R)(1-e^{-\alpha (T) d})$ is plotted 
in Fig. S3(b), varying from 70 \% at room temperature to 45 \% at 10 K,
where the reflectivity at the sample surface ($R=0.21$) is assumed to be independent of the temperature.
This effect has been corrected for the data points in Fig. 4(a) in the main text.

Furthermore, the luminescence signals are partially lost when the entrance slit of the monochromator is narrow. 
We considered this effect by calculating the fraction of the luminescence photons entering the monochromator using
$h(T)={\rm erf} (s/M/L)=2/\sqrt{\pi}\int_{0}^{s/M/L} \exp (-x^2) dx$,  
where $s$ is the width of the monochromator slit and $M=5/3$ is the magnification factor. We assumed a temperature-dependent
Gaussian broadening with the diffusion length $L=\sqrt{D_{\rm eff} \tau_{\rm net}}$, whose temperature dependence is
plotted in Fig. S3(c). 
As can be seen in Fig. S3(d), the effect of $h(T)$ is not significant above 10 K and when the slit is open at 0.1 mm.

To summarize, the total PL intensity shown in Fig. 4(a) is 
the spectrally integrated intensity corrected by the factors $g(T)$ and $h(T)$.
It should be noted that these counts are not absolute photon numbers, as we did not consider the efficiencies resulting from 
the solid angle of detection and conversion from photon to electron counts in the CCD camera. 
Thus, the counts are shown in a relative unit on the right axis of Fig. 4(a). 

\begin{figure}[tb]
\centering
\includegraphics[width=8.5cm,clip,bb=15 30 785 543]{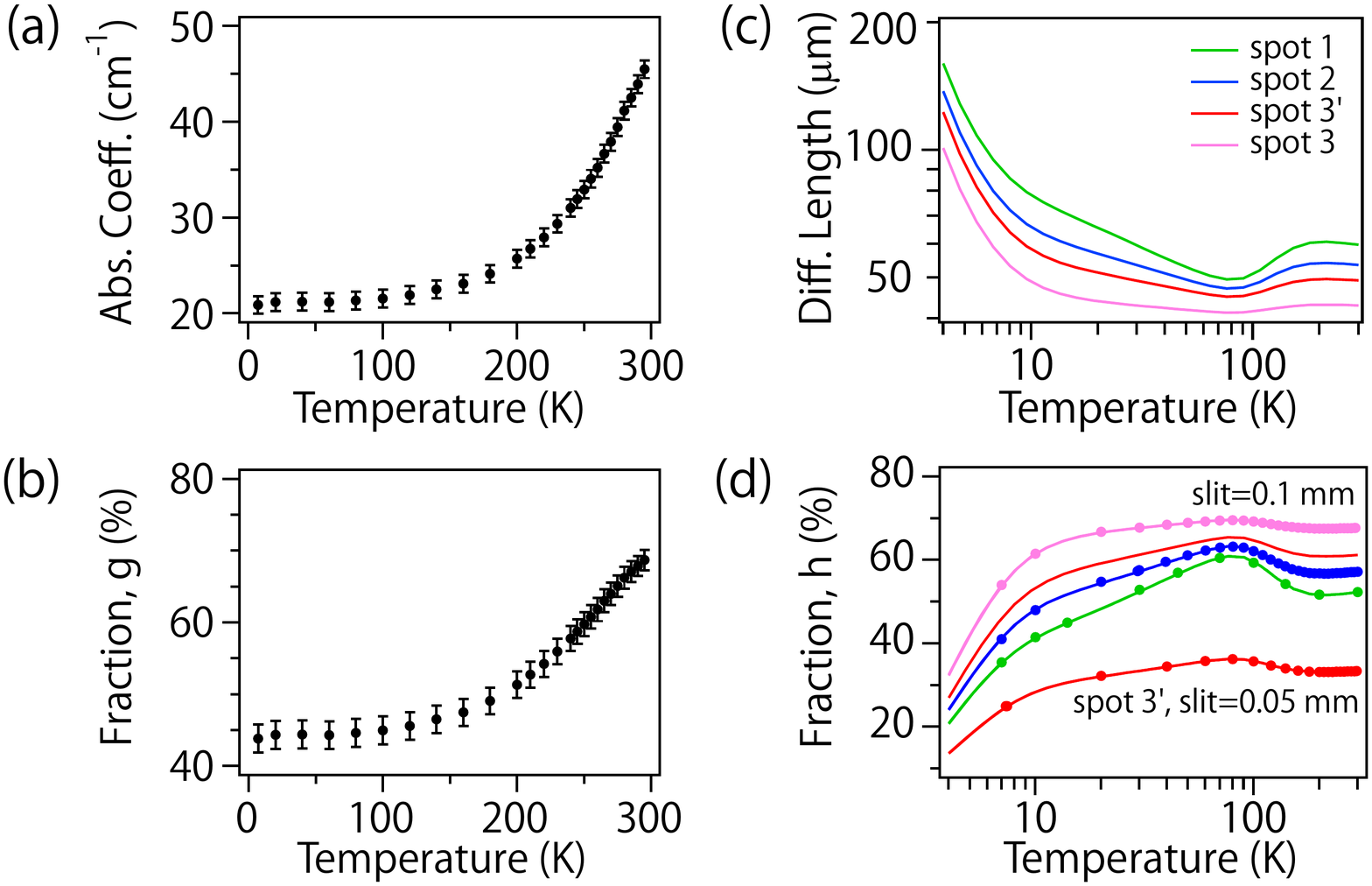}
\caption{(a) Absorption coefficient measured on sample 1 as a function of temperature.
(b) Temperature dependence of the fraction of the laser power absorbed in a 500 $\mu$m-thick sample. 
(c) Temperature dependence of the diffusion length. (d) The fraction $h(T)$ of photons
entering the monochromator after spatial expansion by diffusion.}
\end{figure} 

\section*{REFERENCES}
\noindent
[1] H. Morimoto, Y. Hazama, K. Tanaka, and N. Naka, 

Phys. Rev. B {\bf 92}, 201202(R) (2015).

\noindent
[2] Y. Hazama, N. Naka, M. Kuwata-Gonokami, and K. 

Tanaka,
Euro Phys. Lett. {\bf 104}, 47012 (2013).

\end{document}